\documentclass{aa}  

\usepackage{graphicx}
\usepackage{txfonts}
\usepackage{booktabs}
\usepackage{natbib}

\usepackage{hyperref}
\hypersetup{colorlinks,allcolors=black}

\usepackage[normalem]{ulem}
\usepackage{siunitx}

\def\Mjup{\hbox{$\mathrm{M}_{\rm Jup}$}}
\def\Rjup{\hbox{$\mathrm{R}_{\rm Jup}$}}

\def\hd20{HD~209458\,b}
\def\gj34{GJ~3470\,b}
\def\hat11{HAT-P-11\,b}

\def\hep{He$^{+}$}

\def\het{He~(2$^{3}$S)}
\def\mlr{$\dot M$}
\def\lya{Ly$\alpha$}
\def\earth{\oplus}
\def\halpha{H$\alpha$ }
\def\gs{g\,s$^{-1}$}

\def\rp{$R_{\rm P}$}

\newcommand{\tstar}{GJ~3470\xspace}
\newcommand{\tplanet}{GJ~3470\,b\xspace}

\begin{document} 

   \title{A He {\sc i} upper atmosphere around the warm Neptune GJ~3470\,b}
   \titlerunning{Helium in GJ~3470\,b}
   %\subtitle{}

%----Main-----
  \author{E. Palle\inst{\ref{iiac},\ref{iull}}
  \and L.~Nortmann\inst{\ref{iiac},\ref{iull}}
  \and N.~Casasayas-Barris\inst{\ref{iiac},\ref{iull}}
%----WG paper-----
  \and M.~Lamp\'on\inst{\ref{inst:iaa}} 
  \and M.~L\'opez-Puertas\inst{\ref{inst:iaa}}
  \and J.\,A.~Caballero\inst{\ref{inst:cab}}
  \and J.~Sanz-Forcada\inst{\ref{inst:cab}} 
  \and L.\,M.~Lara\inst{\ref{inst:iaa}}  
  \and E.~Nagel\inst{\ref{inst:ham},\ref{inst:tls}} 
  \and F.~Yan\inst{\ref{inst:gott}} 
%----Alphabet-----
  \and F.~J.~Alonso-Floriano\inst{\ref{inst:leiden}}
  \and P.\,J.~Amado\inst{\ref{inst:iaa}}
  \and G.~Chen\inst{\ref{iiac},\ref{iull},\ref{inst:purp}}
  \and C.~Cifuentes\inst{\ref{inst:cab}} 
  \and M.~Cort\'es-Contreras\inst{\ref{inst:cab}}
  \and S.~Czesla\inst{\ref{inst:ham}} 
%  \and S.~Khalafinejad\inst{\ref{inst:lsw}}
  \and K.~Molaverdikhani\inst{\ref{inst:mpia}} 
  \and D. Montes\inst{\ref{inst:ucm}}
  \and V.\,M.~Passegger\inst{\ref{inst:ham}}
  \and A.~Quirrenbach\inst{\ref{inst:lsw}} 
  \and A.~Reiners\inst{\ref{inst:gott}}
  \and I.~Ribas\inst{\ref{inst:ice},\ref{inst:ieec}}
  \and A.~S\'anchez-L\'opez\inst{\ref{inst:iaa}}
  \and A. Schweitzer\inst{\ref{inst:ham}}
  \and M.~Stangret\inst{\ref{iiac},\ref{iull}} 
  \and M.~R.~Zapatero~Osorio\inst{\ref{inst:cab}}
  \and M.~Zechmeister\inst{\ref{inst:gott}} 
 }

%   \author{N.~Casasayas-Barris\inst{1,2} \and E.~Pallé\inst{1,2} \and F.~Yan\inst{3} \and G.~Chen\inst{4} \and R.~Luque\inst{1,2} \and M.~Stangret\inst{1,2} \and E.~Nagel\inst{5} \and M.~Zechmeister\inst{3} \and  M.~Oshagh\inst{3} \and  J.~Sanz-Forcada\inst{6} \and L.~Nortmann\inst{1,2} \and F.~J.~Alonso-Floriano\inst{7} \and P.~J.~Amado\inst{8} \and J.~A.~Caballero\inst{6} \and S.~Czesla\inst{5} \and S.~Khalafinejad\inst{14} \and M.~López-Puertas\inst{8} \and J.~López-Santiago\inst{9,10} \and K.~Molaverdikhani\inst{11} \and D. Montes\inst{15} \and A.~Quirrenbach\inst{14} \and A.~Reiners\inst{3} \and I.~Ribas\inst{12,13} \and A.~Sánchez-López\inst{8} \and M.~R.~Zapatero-Osorio\inst{6}
% }

%   \and et al\inst{\ref{iiac}} 
%   \and P. Montanes Rodriguez\inst{\ref{iiac},\ref{iull}} 
%   \and F. Murgas\inst{\ref{iiac},\ref{iull}} 
%   \and N. Narita\inst{\ref{iutda},\ref{iabc},\ref{ijsta},\ref{inao},\ref{iuteps}} 
%   \and D.~Hidalgo~Soto\inst{\ref{iiac},\ref{iull}}
%   \and V.~Bejar\inst{\ref{iiac},\ref{iull}}
%Aphabetical   
%   \and N.~Casasayas~Barris\inst{\ref{iiac},\ref{iull}}
%   \and N.~Crouzet\inst{\ref{iiac},\ref{iull}}
%   \and A.~Fukui\inst{\ref{iuteps}}
%   \and P.~Klagyivik\inst{\ref{iiac},\ref{iull}}
%   \and N.~Kusakabe \inst{\ref{iabc},\ref{inao}} 
%   \and R.~Luque~Ram\'irez\inst{\ref{iiac},\ref{iull}}
%   \and M.~Mori\inst{\ref{iutda}}
%   \and T.~Nishiumi\inst{\ref{ikyo}}
%   \and J.~Prieto-Arranz\inst{\ref{iiac},\ref{iull}}
%   \and M.~Tamura\inst{\ref{iutda},\ref{iabc},\ref{inao}}
%- TESS Co-authors 
%   }

  \institute{
  	     \label{iiac} Instituto de Astrof\'isica de Canarias (IAC), E-38200 La Laguna, Tenerife, Spain
  	\and 
  	    \label{iull} Deptartamento de Astrof\'isica, Universidad de La Laguna (ULL), E-38206 La Laguna, Tenerife, Spain
	\and 
        \label{inst:iaa} Instituto de Astrof\'isica de Andaluc\'ia (IAA-CSIC), Glorieta de la Astronom\'ia s/n, 18008 Granada, Spain
        \and
            \label{inst:cab} Centro de Astrobiolog\'ia (CSIC-INTA), ESAC, Camino bajo del castillo s/n, 28692 Villanueva de la Ca\~nada, Madrid, Spain
        \and
            \label{inst:ham} Hamburger Sternwarte, Universit\"at Hamburg, Gojenbergsweg 112, 21029 Hamburg, Germany
        \and
            \label{inst:tls} Th\"uringer Landessternwarte Tautenburg, Sternwarte 5, 07778 Tautenburg, Germany
        \and 
            \label{inst:gott} Institut f\"ur Astrophysik, Georg-August-Universit\"at, Friedrich- Hund-Platz 1, 37077 Göttingen, Germany
        \and
            \label{inst:leiden} Leiden Observatory, Leiden University, Postbus 9513, 2300 RA, Leiden, The Netherlands
        \and
            \label{inst:purp} Key Laboratory of Planetary Sciences, Purple Mountain Observatory, Chinese Academy of Sciences, Nanjing 210033, China     
        \and
            \label{inst:mpia} Max-Planck-Institut f\"ur Astronomie, K\"onigstuhl 17, 69117 Heidelberg, Germany
        \and
            \label{inst:ucm} Departamento de F\'isica de la Tierra y Astrof\'isica and IPARCOS-UCM (Intituto de F\'isica de Part\'iculas y del Cosmos de la UCM), Facultad de Ciencias F\'isicas, Universidad Complutense de Madrid, 28040 Madrid, Spain
        \and
            \label{inst:lsw} Landessternwarte, Zentrum f\"ur Astronomie der Universit\"at Heidelberg, K\"onigstuhl 12, 69117 Heidelberg, Germany
        \and
            \label{inst:ice} Institut de Ci\`encies de l'Espai (ICE, CSIC), Campus UAB, c/ de Can Magrans s/n, 08193 Bellaterra, Barcelona, Spain
        \and 
            \label{inst:ieec} Institut d'Estudis Espacials de Catalunya (IEEC), 08034 Barcelona, Spain
  }

  \date{Received 13 February 2020 / Accepted 16 April 2020}

  \abstract
   {High resolution transit spectroscopy has proven to be a reliable technique for the characterization of the chemical composition of exoplanet atmospheres. Taking advantage of the broad spectral coverage of the CARMENES spectrograph, we initiated a survey aimed at characterizing a broad range of planetary systems. Here, we report our observations of three transits of \tplanet with CARMENES in search of \het\ absorption. On one of the nights, the He~{\sc i} region was heavily contaminated by OH$^-$ telluric emission and, thus, it was not useful for our purposes. The remaining two nights had a very different signal-to-noise ratio (S/N) due to weather. They both indicate the presence of \het\ absorption in the transmission spectrum of \tplanet, although a statistically valid detection can only be claimed for the night with higher S/N. For that night, we retrieved a 1.5$\pm$0.3\% absorption depth, translating into a $R_p(\lambda)/R_p = 1.15\pm 0.14$ at this wavelength. Spectro-photometric light curves for this same night also indicate the presence of extra absorption during the planetary transit with a consistent absorption depth. The \het\ absorption is modeled in detail using a radiative transfer code, and the results of our modeling efforts are compared to the observations. We find that the mass-loss rate, \mlr, is confined to a range of 3\,$\times\,10^{10}$\,\gs\ for $T$ = 6000\,K to 10\,$\times\,10^{10}$\,\gs\ for $T$ = 9000\,K. We discuss the physical mechanisms and implications of the He~{\sc i} detection in \tplanet and put it in context as compared to similar detections and non-detections in other Neptune-size planets. We also present improved stellar and planetary parameter determinations based on our visible and near-infrared observations. 
   }
%   {Indeed.}
%   {Indeed.}
%   {Indeed}
%   {}

   \maketitle

\section{Introduction}
\label{sec:introduction}

High resolution spectroscopy has been established over the past few years as a major tool for the characterization of exoplanet atmospheres. The cross-correlation technique of planetary models and observed spectral time series has allowed for the detection of CO, CH$_4$, and H$_2$O molecules in the atmospheres of hot Jupiters \citep{Snellen2010, deKok2013, Birkby2013, Guilluy2019} and holds the key to spectroscopic characterization of rocky worlds with the upcoming extremely large telescopes \citep{Palle2011, Snellen2013}.
 
Moreover, using high resolution transmission spectroscopy, we are not only able to detect chemical species in the atmosphere of exoplanets, but also to resolve their spectral lines. If the signal-to-noise ratio (S/N) % S/N
of the final transmission spectrum is high enough, it is possible to obtain temperature and pressure profiles of the upper atmosphere by adjusting isothermal models to different regions of the lines (from core to wings), whose origins reside in different layers of the atmosphere \citep{Wy2015, Wy2017, Casasayas2018}. 

The ability to measure and track line profiles can greatly help in the study of atmospheric escape, which is an important process for understanding planetary physical and chemical evolution. In the past, studies of atmospheric escape relied mostly on space-based observations of the hydrogen Ly$\alpha$ line in the far ultraviolet \citep{Vidal2003}, a spectral region with limited access and strongly affected by interstellar absorption. 

However, the near-infrared coverage of spectrographs such as CARMENES and GIANO gives access to poorly-explored exoplanet atmospheric features, including the triplet line feature of metastable neutral helium at 10830\,{\AA}. 
This line was proposed as a tracer for atmospheric evaporation in general by \citet{Seager2000} and for particular targets by \citet{Oklopcic2018}. 
In this process, intense high-energy irradiation from the host star causes the atmosphere of a hot gas planet to continuously expand resulting in mass flowing away from the planet \citep{Lammer2013,Lundkvist2016}. 
With the recent detections of He~{\sc i} with low \citep{Spake2018} and high resolution spectroscopy \citep{Nortmann2018Science, Allart2018, Salz2018He}, it has been proven that this line is a powerful tool for studying the extended atmospheres, mass-loss, and winds in the upper-atmospheres, and for tracking the possible presence of cometary-like atmospheric tails.

Atmospheric erosion by high-energy stellar radiation is believed to play a major role in shaping the distribution of planet radii. Planets with H/He-rich envelopes can be strongly evaporated by stellar irradiation. The evaporation theory predicts the existence of an ``evaporation valley'' with a paucity of planets at $\sim 1.7\,\mathrm{R_{\earth}}$ \citep{Seager2000, Owen2013}. The radius distribution of small planets ($R_p < 4.0\,\mathrm{R_{\earth}}$) is bi-modal; small planets tend to have radii of either $\sim 1.3\,\mathrm{R_{\earth}}$ (super-Earths) or $\sim 2.6\,\mathrm{R_{\earth}}$ (sub-Neptunes), with a dearth of planets at $\sim 1.7\,\mathrm{R_{\oplus}}$ \citep{Fulton17, van2017, Fulton18}. This gap suggests that all small planets might have solid cores, while the cores of sub-Neptune planets are expected to be surrounded by H/He-rich envelopes that significantly enlarge the planetary radii as they are optically thick, while accounting for only 1\,\% of the total planetary mass. Terrestrial cores can also be surrounded by a thin atmosphere or possess no atmosphere at all, making up the population of super-Earths centered at $R_p \sim 1.3\,\mathrm{R_{\oplus}}$

\begin{table*}
\centering
\caption{Observing log of the \tplanet transit observations. RV is the averaged barycentric Earth radial velocity during the night.}
\begin{tabular}{ccc ccc ccc}
    \hline
    \hline
    \smallskip
Night $t$ & Date  & Start UT & End UT  &  $t_\mathrm{exp}$ [s] & $N_\mathrm{obs}$ & Airmass & S/N & RV [km/s] \\
    \smallskip
    \smallskip
1 & 2018 Dec 16 & 22:23 & 02:05 & 498 & 23 & 1.85$\rightarrow$1.08$\rightarrow$1.08 & 26 & 7.15 \\ 
%\hline
%\\[-1em]
2 & 2018 Dec 26 & 21:38 & 03:13 & 498 & 34 & 1.9$\rightarrow$1.079$\rightarrow$1.136 & 66 & 12.24\\
%\hline
%\\[-1em]
3 & 2019 Jan 05 & 21:54 & 03:27 & 498 & 35 & 1.48$\rightarrow$1.078$\rightarrow$1.25 & 61 & 16.90\\
\\[-1em]
\hline
%\hline
\end{tabular}\\
\label{Tab:Obs}
\end{table*}
	
\tplanet \citep{2012A&A...546A..27B} is a warm Neptune	
%($R = 0.37\pm 0.05$\,\Rjup, $M = 0.039\pm 0.004$\,\Mjup), 
($R = 3.88\pm 0.32$\,$R_{\earth}$, $M = 12.58\pm 1.3$\,$M_{\earth}$),
with an equilibrium temperature of 547\,K and a period of 3.33\,d, located very close to the Neptunian desert. 
Previous atmospheric studies have inferred a hazy, low-methane or metal-rich atmosphere from {\em Hubble Space Telescope} observations \citep{Ere14} and a Rayleigh slope in the visible range \citep{Nascimbeni13,Chen17}. 
While Earth-size and super-Earth planets still remain out of the reach of current instrumental capabilities for evaporation studies, \tplanet is an excellent target for study of such processes. 
Indeed, \citet{Bou18} already reported the existence of a giant hydrogen exosphere around GJ~3470b and derived a high mass-loss rate. 
Here we present observations of this target in search for the absorption features of the \het\ triplet. During the process of writing and refereeing of this manuscript, a similar independent work was reported by \citet{Ninan2019}.

\section{Observations and data analysis}

\subsection{CARMENES spectroscopy}
\label{sec:obs}

The transit of \tplanet was observed three times with the CARMENES spectrograph  \citep{CARMENES, CARMENES18} at the Calar Alto Observatory, on the nights of 16 and 26 December 2018, and on 5 January 2019 (nights 1, 2, and 3, hereafter). CARMENES covers simultaneously the visual ($0.52$--$\SI{0.96}{\micro\metre}$) and near-infrared ($0.96$--$\SI{1.71}{\micro\metre}$) spectral ranges with its two channels. A log of the observations, including start and ending times, airmass intervals, and S/Ns can be found in Table~\ref{Tab:Obs}. 
Altogether, we collected 13, 14, and 13 in-transit spectra on each night, respectively, using the criteria that at least half the exposure time was taken inside the first and fourth contact interval. 
Following the same criteria, we also obtained 10, 20, and 22 out-of-transit spectra on nights 1, 2, and 3, respectively.

During the observations, fiber A was fed by the light of the GJ~3470 star and fiber B felt on the sky at about 1.5\,arcmin to the target. The spectra of both fibers were extracted from the raw frames using the CARACAL pipeline \citep{SERVAL}. In the standard data flow \citep{2016SPIE.9910E..0EC}, fiber A spectra are extracted using flat optimized extraction while fiber B spectra are extracted with a simple aperture. Here, we also extracted fiber B with flat optimized extraction so that the spectra of both fibers underwent the same processing scheme.

\begin{table}
\centering
\caption{Stellar parameters of GJ~3470.}
\label{tbl:star}
\begin{tabular}{l c r} 
    \hline
    \hline
    \noalign{\smallskip}
Parameter           & Value             & Reference \\		 
    \noalign{\smallskip}
    \hline
    \noalign{\smallskip}
    \multicolumn{3}{c}{\em Name and identifiers} \\
    \noalign{\smallskip}
Name                & LP 424--4         & Luy79 \\
GJ                  & 3470              & GJ91 \\
Karmn               & J07590+153        & AF15 \\
    \noalign{\smallskip}
    \multicolumn{3}{c}{\em Key parameters} \\
    \noalign{\smallskip}
$\alpha$            & 07:59:05.84       & {\em Gaia} DR2 \\
$\delta$            & +15:23:29.2       & {\em Gaia} DR2 \\
$G$ (mag)           & 11.3537$\pm$0.0013 & {\em Gaia} DR2 \\
$J$ (mag)           & 8.794$\pm$0.026 & 2MASS \\
Spectral type       & M2.0\,V           & Lep13 \\
    \noalign{\smallskip}
    \multicolumn{3}{c}{\em Parallax and kinematics} \\
    \noalign{\smallskip}
$\pi$ (mas)         & 33.96$\pm$0.06    & {\em Gaia} DR2 \\
$d$ (pc)            & 29.45$\pm$0.05    & {\em Gaia} DR2 \\
$\mu_\alpha \cos{\delta}$ (mas\,a$^{-1}$) & --185.73$\pm$0.11 & {\em Gaia} DR2 \\
$\mu_\delta$ (mas\,a$^{-1}$) & --57.26$\pm$0.06 & {\em Gaia} DR2 \\
$V_r$ (km\,s$^{-1}$)$^{a}$ & +26.5169$\pm$0.0005 & Bou18 \\
$U$ (km\,s$^{-1}$)  & --32.04$\pm$0.21   & This work \\
$V$ (km\,s$^{-1}$)  & --12.42$\pm$0.10   & This work \\
$W$ (km\,s$^{-1}$)  & --15.37$\pm$0.10   & This work \\
Kinematic population & Young disc       & This work \\
    \noalign{\smallskip}
    \multicolumn{3}{c}{\em Photospheric parameters} \\
    \noalign{\smallskip}
$T_{\rm eff}$ (K)   & 3725$\pm$54      & This work \\ % Gaia 3775.50
$\log{g}$           & 4.65$\pm$0.06       & This work \\
{[Fe/H]}            & +0.420$\pm$0.019         & This work \\ % Santos: 0.08
$v \sin{i}$ (km\,s$^{-1}$) & $\lesssim$2 & Bon12 \\
    \noalign{\smallskip}
    \multicolumn{3}{c}{\em Physical parameters} \\
    \noalign{\smallskip}
$L$ (10$^{-4}$\,$L_\odot$) & 390$\pm$5 & This work \\
$R$ ($R_\odot$)     & 0.474$\pm$0.014    & This work \\
$M$ ($M_\odot$)     & 0.476$\pm$0.019    & This work \\
Age (Ga)            & 0.6--3.0           & This work \\ 
    \noalign{\smallskip}
    \multicolumn{3}{c}{\em Other parameters} \\
    \noalign{\smallskip}
$P_{\rm rot}$ (d)   & 20.70$\pm$0.15    & Bid15 \\
pEW(H$\alpha$) (\AA)      & +0.39$\pm$0.09 & Gai14 \\
$\log{R'_{\rm HK}}$ & --4.91$\pm$0.11   & SM15  \\
$F_{\rm 5-100\,\AA}$ (10$^{27}$\,erg\,s$^{-1}$)   & 2.3   & Bou18 \\
$F_{\rm 100-504\,\AA}$ (10$^{27}$\,erg\,s$^{-1}$)   & 2.7   & Bou18 \\    
\noalign{\smallskip}
\hline
\end{tabular}   
\begin{list}{}{}
\item {\bf References.} 
AF15: \citet{2015A&A...577A.128A};
Bid14: \citet{Biddle2014};
Bon12: \citet{2012A&A...546A..27B};
Bou18: \citet{Bou18};
Gai14: \citet{2014MNRAS.443.2561G};
GJ91: \citet{1991adc..rept.....G};
L\'ep13: \citet{2013AJ....145..102L};
Luy79: \citet{1979nlcs.book.....L};
%San13: Santos et al. (2013);
SM15: \citet{2015MNRAS.452.2745S};
2MASS: \citet{2006AJ....131.1163S};
{\em Gaia} DR2: \citet{2018A&A...616A...1G}.
{\bf Notes.} $^{a}$ Soubiran et al. (2018) tabulated $V_r$ = +26.341$\pm$0.004\,km\,s$^{-1}$, but their uncertainties did not include gravitational redshift or photospheric convection.
%\item[$^{a}$] The \teff estimate is based on \citet{Rajpurohit2013} and \citet{Pecaut2013}, the $R_\star$ estimate on \citet{Schweitzer2019}, and the $M_\star$ estimate on \citet{Maldonado2015}.
\end{list}
\end{table}

\subsection{Target star parameters}
%\subsection{Target star parameters retrieval}

The star GJ~3470 was first cataloged as a high proper motion star in the Luyten-Palomar survey \citep{1979nlcs.book.....L}.
It went almost unnoticed until \citet{2012A&A...546A..27B} discovered the transiting planet around it.
Since then, and especially with the advent of {\em Gaia} \citep{2018A&A...616A...1G}, the stellar parameters of GJ~3470 have been better measured.

In Table~\ref{tbl:star} we compile a comprehensive list of stellar parameters of \tstar, either from the literature or derived by us.
When there are different published parameter determinations (e.g., spectral type, proper motion), we list the most precise or the most recent ones.
%In particular, we re-determined several of these parameters.

We determined the photospheric parameters $T_{\rm eff}$, $\log{g}$, and [Fe/H] following the methods described by \citet{2019A&A...627A.161P}, using the combined VIS+NIR spectra of the two CARMENES channels.
The physical stellar parameters $L$, $R$, and $M$ were determined following
\citet{2019A&A...625A..68S}, i.e., we measured the luminosity $L$ by using
the {\em Gaia}~DR2 parallax and integrated multi-wavelength photometry from $B$ to $W4$,
applied Stefan-Boltzmann's law to obtain the radius $R$, and, finally, used the linear mass-radius relation from
\citet{2019A&A...625A..68S}
to arrive at the mass $M$.

Our photospheric parameters ($T_{\rm eff}$, $\log{g}$, and [Fe/H]) are consistent with \citet{Demory2013}. Their mass was based on the empirical mass-magnitude relation of \citet{2000A&A...364..217D} and, hence, it differs by the same amount from our value as results from \citet{2000A&A...364..217D} differ from the updated mass-magnitude relation of \citet{2019ApJ...871...63M}. Our method, however, agrees very well with the updated mass-magnitude relation \citep[c.f.,][]{2019A&A...625A..68S}. The radii determination of \citet{Demory2013} or \citet{2016MNRAS.463.2574A}, however, were based on the average density inside the planetary orbit, which added an additional uncertainty.

In addition, we also used the latest astrometric and absolute radial velocity data of {\em Gaia} for determining Galactocentric space velocities $UVW$ and assigning GJ~3470 to the Galactic young disc population.
%As \citet{2012A&A...538A.151S}, we failed to assign membership of GJ~3470 to the Hyades supercluster \citep{1990PASP..102..166E}.
We estimated a stellar age between 0.6\,Ga and 3.0\,Ga, which is consistent with its kinematic population, the presence of H$\alpha$ in absorption (in spite of its M2.0\,V spectral type), the faint Ca~{\sc ii} H\&K emission, the relatively slow rotation (quantified by the low rotational velocity and long rotational period), and its weak X-ray emission, as well as with previous determinations in the literature \citep[e.g.,][]{Bou18,2012A&A...546A..27B}. 

We also searched the literature for additional information on GJ~3470.
What was of particular interest is the $i'$- and $z'$-band lucky imaging of the star by \citet{2015A&A...579A.129W}, who derived upper limits on the existence of targets brighter than $\Delta z' \approx$ 4.0\,mag and 6.0\,mag at 0.25\,arcsec and 5\,arcsec, respectively.
These limits translated into the absence of objects at the substellar boundary at separations beyond 150\,au and more massive than 0.1\,M$_\odot$ down to 7\,au, approximately.

For the system parameters, throughout the rest of the paper, we adopt the stellar velocity semi-amplitude $K_{\rm star}$ from \citet{2012A&A...546A..27B}. For the planet parameters, we recalculated here the radius, mass, density, and equilibrium temperature values (see Table~\ref{Tab:comp}) based on the stellar parameters of Table~\ref{tbl:star}. The remaining values were taken from \citet{Bou18} and references therein. We calculated the velocity semi-amplitude of the planet $K_{\rm planet}$ from these values.

\begin{figure}
	\centering
	\includegraphics[width=\columnwidth]{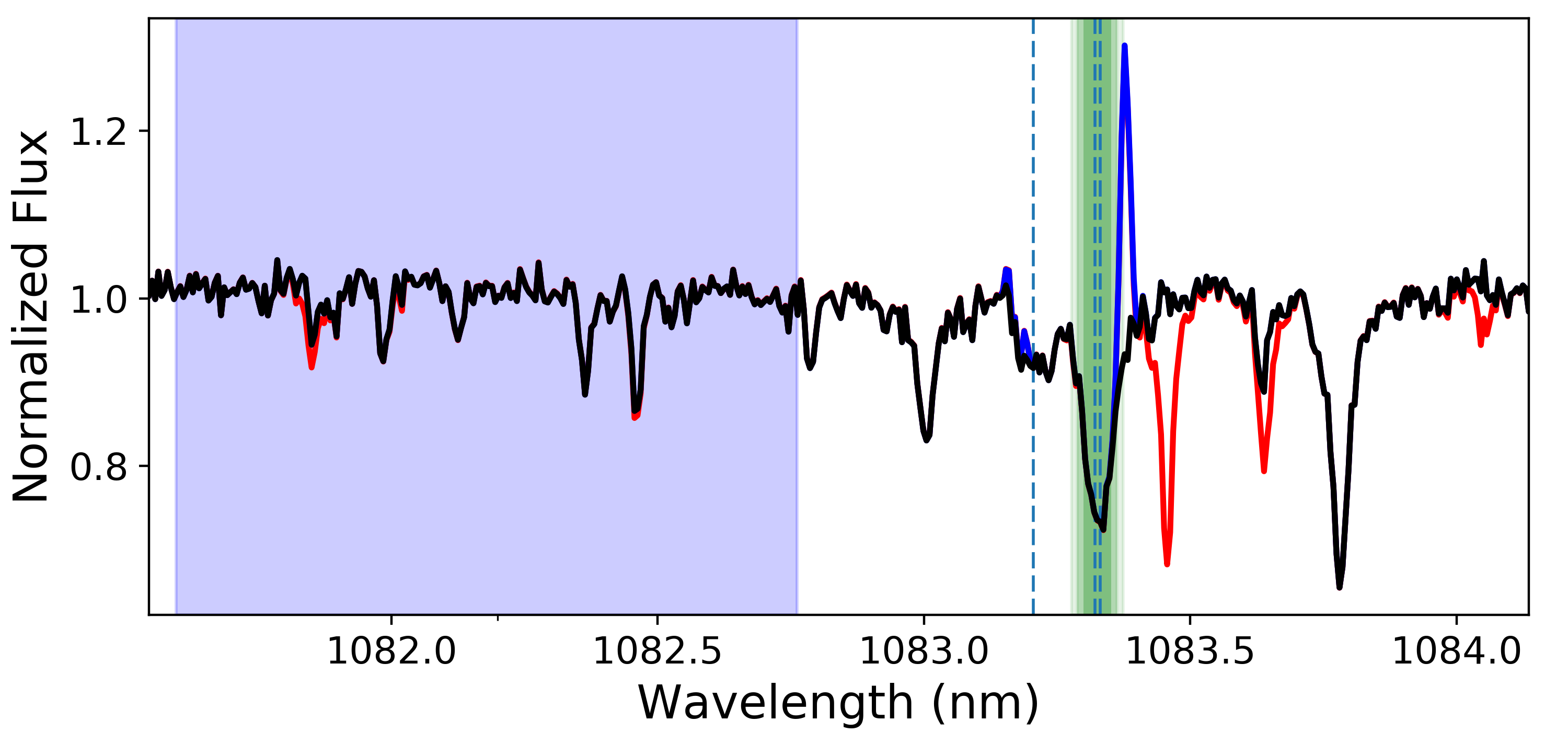}%norm_further_away_1Abin.png}
	\caption{Zoom of one CARMENES spectrum of GJ~3470 in the wavelength region containing the He~{\sc i} triplet. In red the raw spectra after standard data reduction is plotted. Over-plotted in blue is the same spectrum after removal of the telluric features (mainly water in this region) using {\tt molecfit}. In black is the same spectrum after adjusting and removing also the OH$^-$ spectral features. In the figure, the wavelength region used to normalize the continuum of all spectra is marked with a blue shade, and the region around the He~{\sc i} line cores used to calculate the spectro-photometric transit light curves is marked with a shaded green region. }
	\label{fig:carm_spec}
\end{figure}

%-===========================================================================
%\section{Methods}
%\label{sec:methods}

\begin{figure*}
	\centering
	\includegraphics[width=\columnwidth]{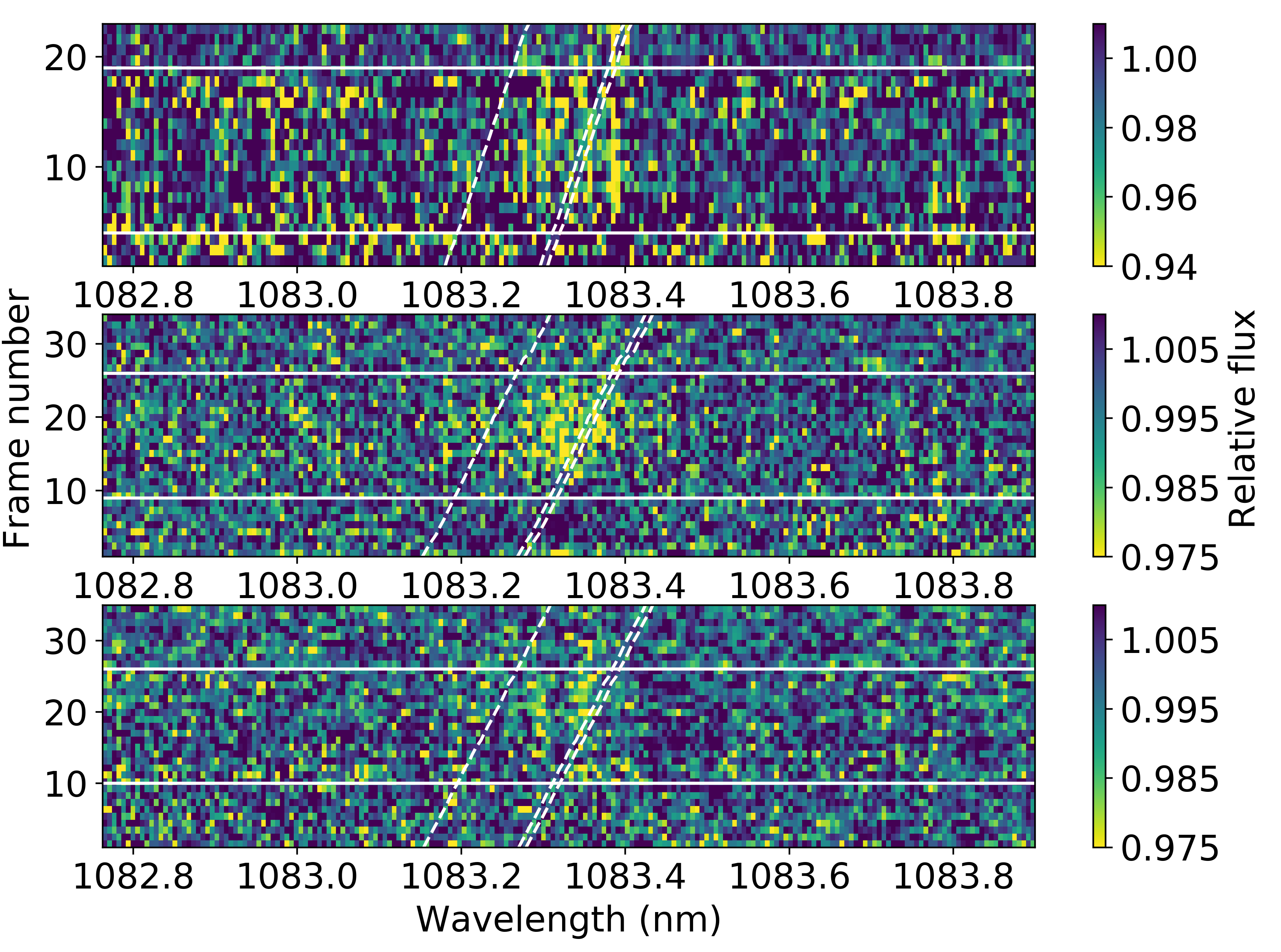}
	\includegraphics[width=\columnwidth]{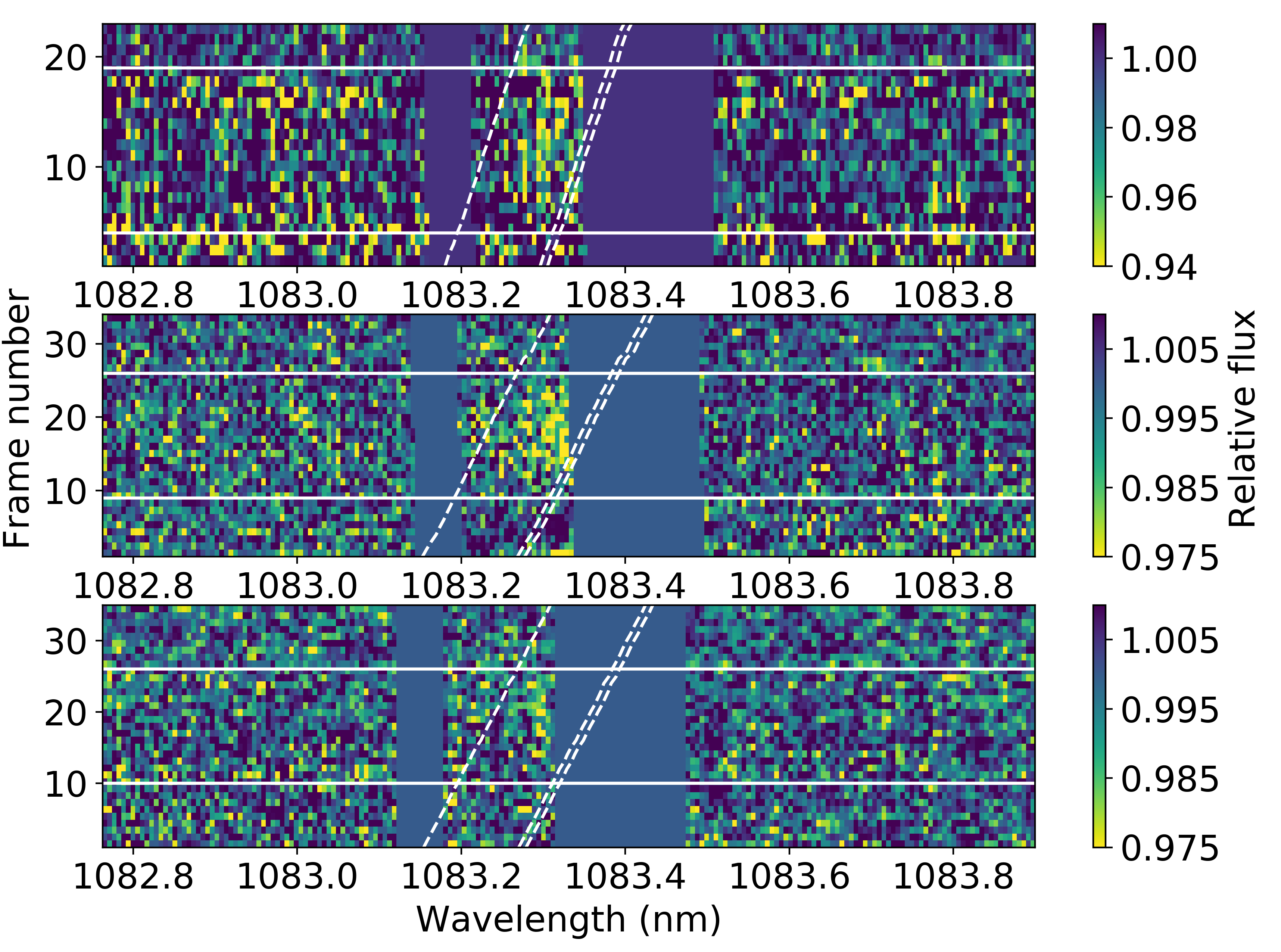}
	\caption{Observed 2D residual maps after dividing each spectrum by the master-out spectrum. Form top to bottom are nights 1, 2 and 3, respectively. The data on the right and left panels are exactly the same, but in the right panel, the regions affected by OH$^-$ contamination are masked to illustrate the amount of usable data for each night. The maps comprise the region around the He~{\sc i} triplet, and are shown in the stellar rest frame. The horizontal white bars mark the beginning (T1) and end (T4) of the transit. The tilted dashed lines mark the expected planetary trail of triplet. Note the different color scale between night 1 and nights 2 and 3. }
	\label{fig:residualmaps}
\end{figure*}

\subsection{Telluric absorption removal}

%The He line is affected by telluric water absorption and OH$^-$ emission lines. We use molecfit to correct the water absorption in the entire spectral region covered by CARMENES (REF TO MOLECFIT).{\color{red}  This is done as described in \citet{Nortmann2018Science} but using the latest version of HITRAN ??(Ask Evangelos to verify). EMAIL EVANGELOS}. The effect   of telluric lines removal are illustrated in Fig.~\ref{fig:carm_spec}.

The He~{\sc i} $\lambda$10830\,{\AA} triplet is contaminated by telluric absorption from atmospheric water vapor and OH$^{-}$ emission \citep{Nortmann2018Science, Salz2018He}. Due to the Earth's barycentric velocity, the relative position between the He~{\sc i} and the telluric features varies with date. To detect the weak planetary signals in the spectral time series, the telluric contribution needs to be removed from the spectra. 

The water vapor removal in each individual spectrum was performed with
\url{molecfit}, which fits synthetic transmission models to the
observations \citep{Smette2015, Kausch2015}.  To adapt the telluric model to the spectra, \url{molecfit} allows the user to convolve the model with an instrumental profile. We analyzed several thousand lines from hollow cathode lamp spectra, which are regularly used as calibration sources, and measured the Gaussian and Lorentzian FWHM components. Based on our analysis, we adopted a value of 5.26\,pixel and 0.75\,pixel for the Gaussian and the Lorentzian FWHM components, respectively. The  determination of the instrumental line spread function is described in more detail in Nagel et al. (submitted). The effect of telluric line removal is illustrated in Figure~\ref{fig:carm_spec}.

The He~{\sc i} triplet lines were also located between OH$^-$ emission lines (see also Fig.~\ref{fig:carm_spec}), which are not accounted for by {\tt molecfit}. 
These lines were also observed in the spectra obtained from fiber B, which was pointed at the sky. 
We corrected the emission lines in fiber A by first modeling the lines in fiber B and then subtracting the model from the spectra of fiber A. 
In fiber B there was no detectable contamination from the stellar spectra. 
To construct the model, we first obtained a master spectrum for fiber B, calculated by summing up all fiber B spectra for a given night. 
To this spectrum, we fitted a Voigt profile to the fiber B OH$^-$ line redwards of the stellar He~{\sc i} lines, and two Gaussian profiles (with the same amplitude and width) to the two weakest OH$^-$ lines bluewards of the stellar He~{\sc i} lines. 
The amplitude of the fit to the strongest OH$^-$ emission could vary for every fiber B spectrum independently, but we kept the values for the positions, widths, and amplitude ratios between strong and weak OH$^-$ lines fixed for all the spectra of a given night.
When allowing the widths of the lines to vary, we found no statistically significant differences in the final results. Finally, when subtracting the model fit of fiber B from fiber A, we applied a scaling factor ($0.88\pm 0.05$) to the model to compensate for the efficiency differences between the two fibers. This factor was calculated from a high S/N observation with CARMENES, and was fixed for all spectra and nights. The error of this factor had no significant impact on our results compared to the standard deviation.

%\subsection{Constructing the transmission spectra}

%\subsection{Spectro-photometric light curves}

%\subsection{ALFOSC Spectrum}
%\label{sec:observations.alfosc}

\section{Results}
\label{sec:results}

\subsection{He {\sc i} transmission spectra}

After correction of the telluric absorption and emission, we normalized all spectra by the mean value of the region between 10815.962\,{\AA} and 10827.624\,{\AA} in vacuum. This region, which lies blue-wards of the He {\sc i} lines, was almost unaffected by telluric absorption and, therefore, gave a robust reference level for the pseudo-continuum (see Fig.~\ref{fig:carm_spec}).

After normalization we aligned all the spectra to the stellar rest frame. %{\color{red} NURIA:Which Ks value do you use? are you calculating the Ks or using a measurement from previous studies?} 
We then calculated a master out-of-transit spectrum by computing the mean spectrum of all spectra obtained out of transit, and divided each individual spectrum (in and out) by this master. This technique has been previously applied in several works \citep{Wy2015, Salz2018He, Casasayas2018}. After removal of the stellar signal, the residual spectra should contain the possible atmospheric planetary signal that, in the stellar rest frame, moves through wavelength space as time progresses, from blue-shifted at the beginning of the transit to red-shifted towards the end. To obtain the transmission spectrum, we aligned these residual spectra to the planet rest frame  %{\color{red} NURIA:Which Kp do you use? is this calculated? Do you consider eccentricity =0?}
and calculated the mean in-transit spectrum between the second and third contacts.

In Figure~\ref{fig:residualmaps} (left panels) the residual maps around the He~{\sc i} triplet are shown for each of the three nights. Also plotted are the ingress start time (first contact) and egress end time (fourth contact), as well as the expected residual trace of a possible planetary signal. Significant positive residual indicative of He~{\sc i} absorption was visible for night 2, but not for the other two nights. The lack of reproducible results could be indicative of a spurious signal. However, it was not clear that this was the case. For night 1, the S/N of the measurements was low due to weather conditions (see Table~\ref{Tab:Obs}), and while a positive signal was also seen at the expected He~{\sc i} wavelengths, the data quality did not allow the signal to reach the $3 \sigma$ significance level required to claim a detection.

For night 3, which had a S/N as night 2, the problem was the contamination of the He~{\sc i} signal by telluric OH$^-$ lines. During the night, the position of the OH$^-$ lines with respect to the He lines changed. To illustrate this, we plot in Figure~\ref{fig:residualmaps} (right panels) the same residual maps. In this figure, however, the spectral regions that during the transit (T1-T4) were overlapping with OH$^-$ lines were masked. It is readily appreciable from the figure that night 3 had the largest OH$^-$ contamination, with practically no unaffected signal from the planet. Thus, we kept the analysis of that night for completeness, but a He~{\sc i} signal detection was not expected for that night even if the planetary absorption was there.

\begin{figure}
	\centering
	\includegraphics[width=\columnwidth]{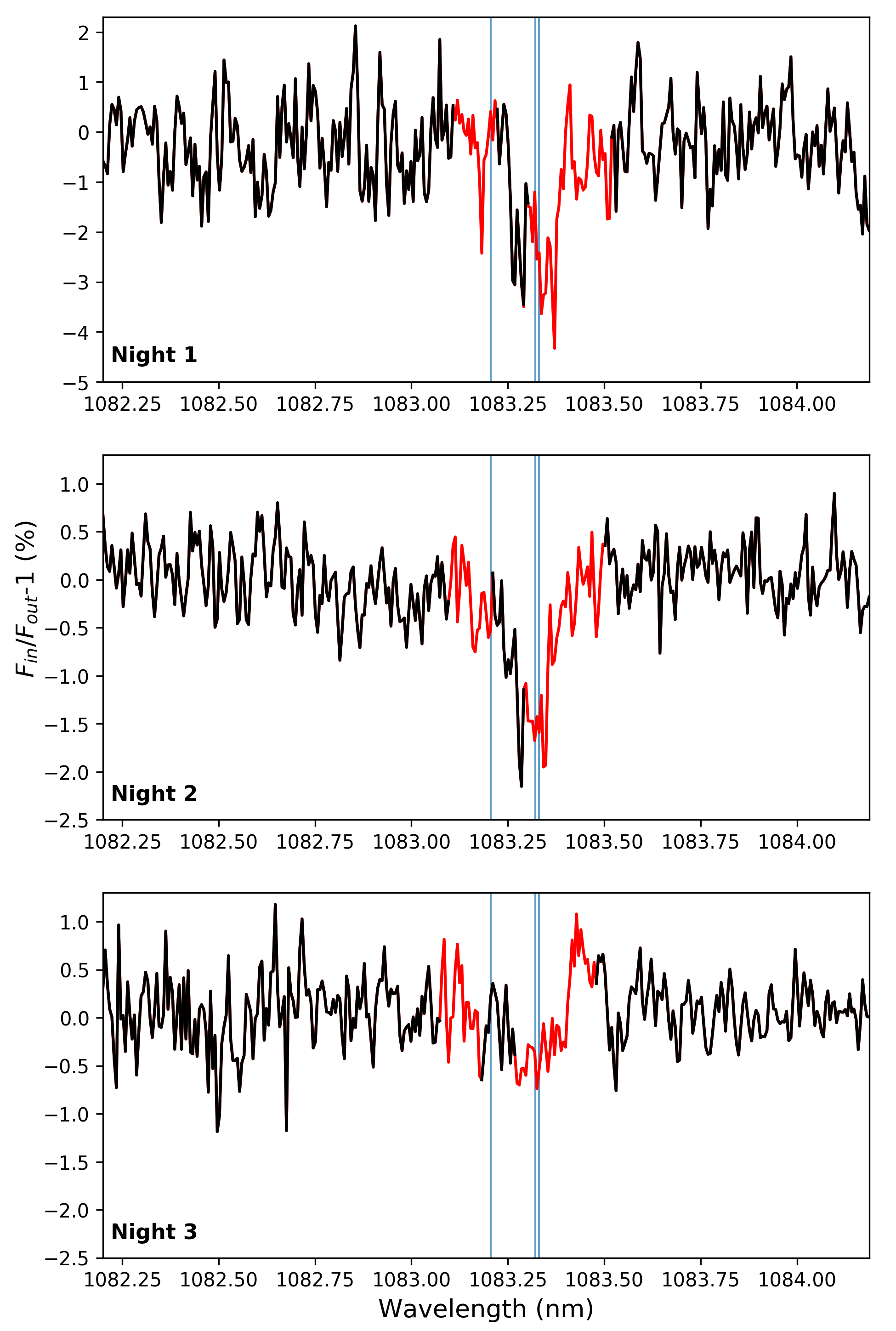}
	\caption{Mid-transit (T2-T3) transmission spectrum around the HeI triplet for night 1, 2 and 3, from top to bottom, respectively. The black line shows the spectral regions unaffected by OH$^-$ lines, while the red line marks the spectral regions affected, and corrected for, OH$^-$ emission. The blue vertical lines mark the helium triplet line center positions. Note the different absorption scale between night 1 and nights 2 and 3.}
	\label{fig:transpec}
\end{figure}

Figure~\ref{fig:transpec} shows the transmission spectrum of He~{\sc i} derived for each of the three nights. The transmission spectrum was calculated in two different ways. The first was by simply masking the OH-affected regions of the spectrum. It is plotted in the figure in black, and it is discontinuous in these affected regions. A second way to calculate the spectra was to correct for OH$^-$ contamination, as described in Section~\ref{sec:obs}. These spectra are over-plotted in red and nothing is masked. The corrected and uncorrected spectra were identical in the common regions. 

In summary, we concluded from the figure that both nights 1 and 2 showed strong absorption features centered in the He~{\sc i} $\lambda$\,10830\,\AA\ triplet. While the scatter for night 1 was large due to the low S/N of the observations, the absorption was clear for night 2, reaching 1.5$\pm$0.3\,\%. Following \citet{Nortmann2018Science}, and using the values in Tables~\ref{Tab:Obs} and ~\ref{tbl:star}, this translated into a planetary radius increase of  $R_p(\lambda)/R_p = 1.15\pm 0.14$, or an equivalent scale height of  $\Delta R_p / H_{eq} = 77\pm 9$.

The absorption in night 1 nearly doubled that of night 2, but there were strong residual features in the transmission spectrum, at the few percent level, that were probably associated to low S/N systematics and were affecting the absorption depth. Night 3, represented at the same scale as night 2, did not show any significant absorption feature. The nightly retrieved absorption depths from the transmission spectrum and the transit light curves (see next section) are given in Table ~\ref{Tab:absorp}.

\subsection{Spectro-photometric light curves}

Spectro-photometric light curves from the spectral data were useful to understand if the absorption features had a temporal variability compatible with the planetary transit. Thus, in order to monitor the temporal behavior of the excess He~{\sc i} absorption, we calculated the transit light curves for this line. To do this, we integrated the counts in band-passes of three different widths (0.40, 0.74, and 0.97\,\AA) centered on the two deepest lines of the He~{\sc i} triplet. This integration was done in the planet rest frame. The summation intervals are marked in Figure~\ref{fig:carm_spec}. The methodology that we followed to build the spectro-photometric light curves was described by \citet{Nortmann2018Science} and \citet{Casasayas2019}. For GJ~3470\,b, the Rossiter-McLaughlin effect on the transmission spectrum and photometric light curves is expected to be negligible.

\begin{figure}
	\centering
	\includegraphics[width=\columnwidth]{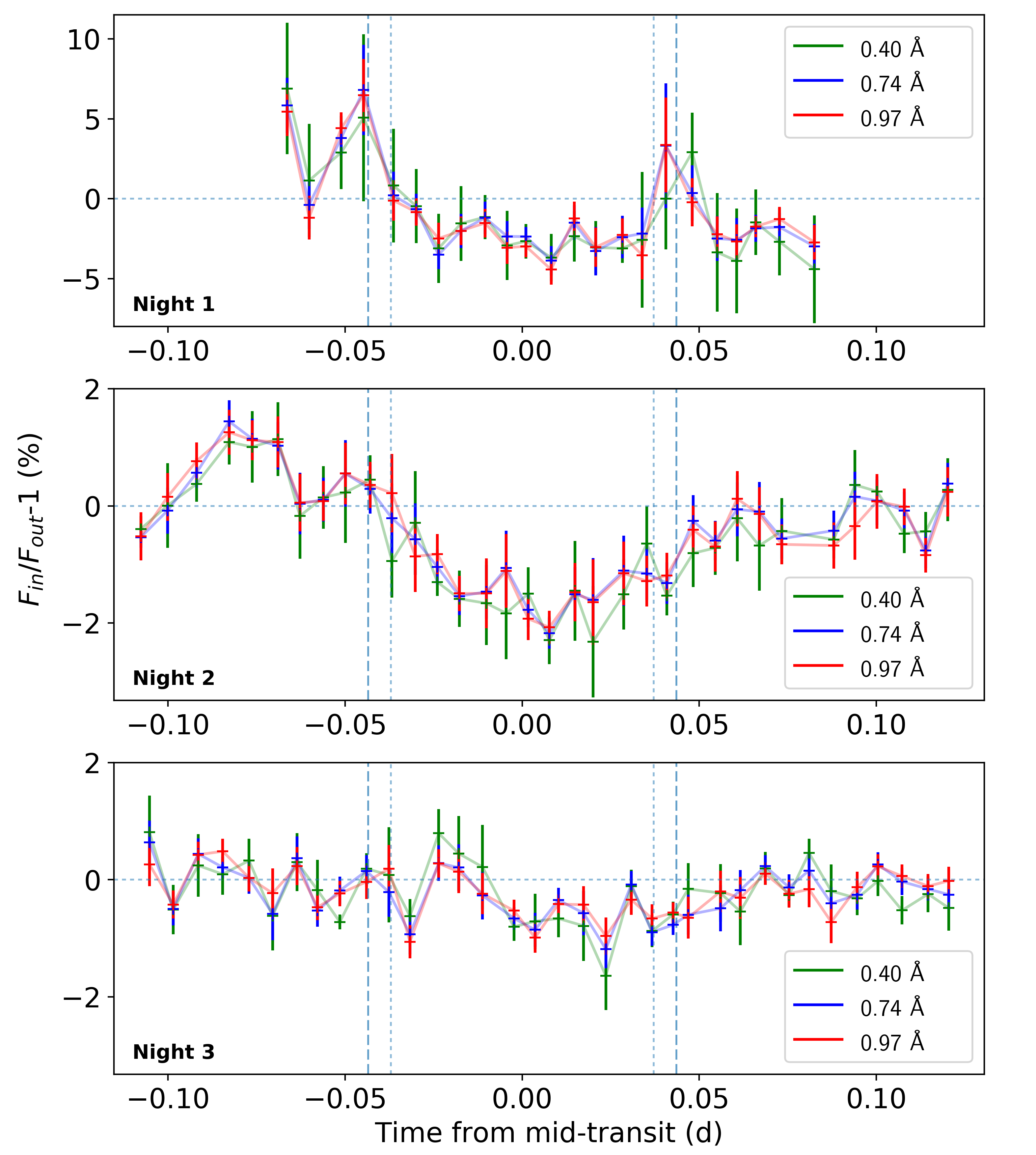}
	\caption{Spectro-photometric light curves of the \het\ absorption of the transit of GJ~3470\,b for each of the three nights: 1 to 3 from top to bottom, respectively. The light curves have been constructed using three different wavelength integration intervals: 0.40\,\AA\ (green), 0.74\,\AA\ (blue), and 0.97\,\AA\ (red). 
	Note the different absorption scale between night 1 and nights 2 and 3.}
	\label{fig:lightcurves}
\end{figure}

In Figure~\ref{fig:lightcurves} we plot the transit light curves for the \het\ absorption for each of the three nights. As in the case of the transmission spectrum, a clear transit was detected only on night 2, while the light curves for nights 1 and 3 were mainly flat. The error bars took into account the individual scatter of each spectrum and the number of points integrated. For night 1, there were a few outliers that coincided in time with ingress and egress, and may resemble of a transit feature, but there was no statistically significant additional absorption during transit. Given the low S/N of the data, this was not surprising as the construction of spectro-photometric light curves from high-dispersion spectroscopy requires a higher S/N \citep{Casasayas2020}. The non-detection of a transit signal for night 3 was consistent with the flat transmission spectrum for the same night.      

For the clear transit of night 2, we observed a transit duration roughy coincident with the expected ingress and egress times. The retrieved depth of the transit was consistent with that retrieved from the transmission spectrum analysis (Table ~\ref{Tab:absorp}). An extra absorption extending further than the egress (tail structure) might be present, but it was not statistically significant within our error bars. New observations minimizing OH$^-$ emission contamination and with larger telescopes will be needed to explore this issue. For night 2 we also observed an ``emission-like'' feature just before the transit, which is already visible in the 2D residual maps in Figure~\ref{fig:residualmaps} as a dark blue region just before the transit start. Currently, we have no explanation for this.  

%\begin{figure}
%	\centering
%	\includegraphics[width=\columnwidth]{LightcurvesNuria.png}
%	\caption{WRITE CAPTION.  ADD RED COLOR on TELLURIC %AFFECTED REGIONS.}
%	\label{fig:lightcurves2}
%\end{figure}

\begin{table}
\centering
\caption{Comparison table of absorption depths retrieved for each individual night$^a$.}
\label{Tab:absorp}
\begin{tabular}{cccc}
    \hline
    \hline
    \smallskip
Night & TS  & TS-Nc & LC \\
    \smallskip
\\[-1em]
\hline
1 & $2.4\pm 0.9$ &  $3.5\pm 0.9$  &  $2.1\pm 0.9$\\ 
2 & $1.5\pm 0.3$  &  $2.2\pm 0.3$ &  $1.4\pm 0.5$\\ 
3 & $0.4\pm 0.2$  &  ...           &  $0.4\pm 0.3$\\ 
\\[-1em]
\hline
%\hline
\end{tabular}\\
\tablefoot{
        \tablefoottext{a}{TS means the value retrieved from the averaged absorption over a 0.4\,{\AA}-wide bin (green shadow in Figure~\ref{fig:carm_spec}). 
        TS-Nc is the same calculation over the transmissions spectrum without accounting for OH-corrected regions (i.e., considering black points only).
        LC refers to the absorption depth retrieved from the transit light curves between second (T2) and third (T3) contacts. 
        For the transmission spectrum the error is simply calculated as the rms over the continuum region 1082.5--1083\,nm.}
    }
\end{table}

\section{Modeling the He~{\sc i} absorption}
\label{sec.modeling}

As previously done in the case of \hd20 \citep{Alonso2019}, we modeled here the \het\ absorption of 
\gj34. Briefly, we used a one-dimensional hydrodynamic and spherically symmetric model together with a non-local thermodynamic equilibrium (non-LTE) model to calculate the \het\ density distribution in the upper atmosphere of the planet \citep{Lampon2020}.
The hydrodynamic equations were solved assuming that the escaping gas has a constant speed of sound, $v_s=\sqrt{k\,T/\mu}$, where $k$ is the Boltzmann constant, $T$ is temperature, and $\mu$ in the mean molecular weight. 
%This approach allows us to decouple the energy budget equation from the mass and momentum equations and then to get an analytical solution. 
This assumption leads to the same analytical solution as the isothermal Parker wind solution. However, the atmosphere is not isothermal. Instead the temperature is such that the $T/\mu$ ratio is constant with altitude, that is, $v_s=\sqrt{k\,T/\mu} = \sqrt{k\,T_0/\bar{\mu}}$, where $\bar{\mu}$ is the average mean molecular weight calculated in the model, and $T_0$ is a model input parameter that is very similar to the maximum of the thermospheric temperature profile calculated by hydrodynamic models that solve the energy balance equation \citep[see, e.g.,][]{Salz2016}. 
The \het\ absorption was later computed by using a radiative transfer code for the standard primary transit geometry \citep{Lampon2020}. The absorption coefficients and wavelengths for the three metastable He~{\sc i} lines were taken from the NIST Atomic Spectra Database\footnote{\tt https://www.nist.gov/pml/atomic-spectra-database}. 
Doppler line shapes were assumed at the atmospheric temperature used in the helium model density, and an additional broadening produced by turbulent velocities was included as described in the reference above. The component of the radial velocity of the gas along the line of sight (towards the observer, i.e., arising from the planet day- and night-sides around the terminator) was also included in order to account for the motion of \het\ as predicted in the hydrodynamic model. From the modeling results, we found that the \het\ distribution is significantly more extended than in the case of \hd20. Hence, we found it necessary to perform the integration of the \het\ absorption up to 10\,\rp.

\begin{figure}
    \includegraphics[angle=90,width=1.0\columnwidth]{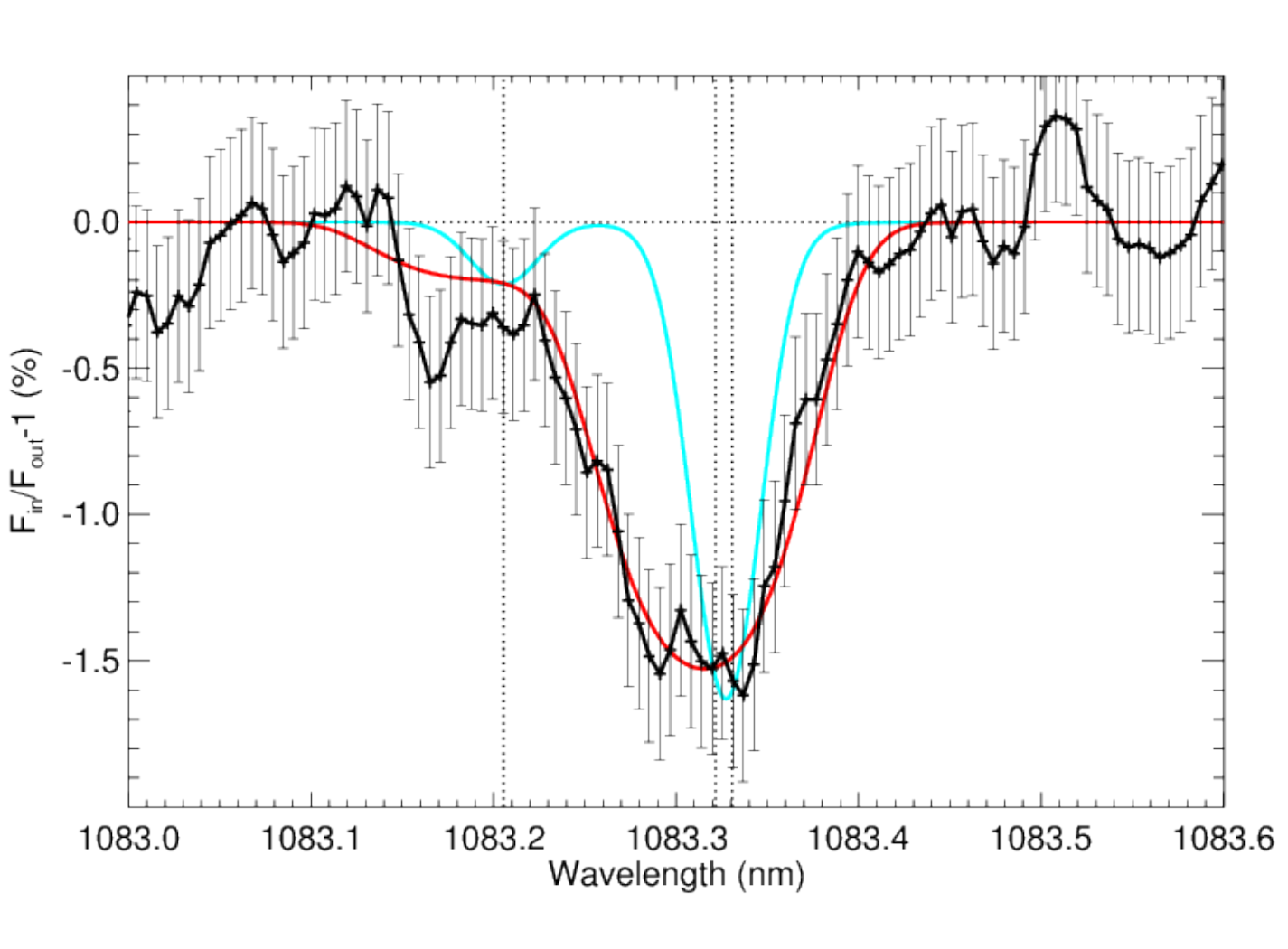}\hspace*{0.35cm}
    \caption{Transmission spectrum of the He~{\sc i} triplet during transit. Measured absorption (+), and their respective estimated errors, are shown in black. The data are the same as in Fig.\,\ref{fig:transpec} but with a three-point running mean applied. The cyan curve shows the absorption profile when only the Doppler and turbulence broadenings are included. The red curve is the best-fit model obtained for an effective temperature of 6000\,K, a mass-loss rate ($\dot{M}$) of 3$\times\,10^{10}$\,\gs\, and a H/He mole-fraction ratio of 90/10. This calculation included, in addition to the Doppler and turbulence broadenings, the broadening induced by the radial velocities of the model and an additional blue net wind of --3.2\,km\,s$^{-1}$. The positions of the three He~{\sc i} lines are marked by vertical dotted lines.}  
    \label{absorption}
\end{figure}

Figure~\ref{absorption} shows the observed transmission spectrum of night 2, together with a calculation performed with the model described above for an effective temperature of 6000\,K and a sub-stellar mass-loss-rate of 3\,$\times\,10^{10}$\,\gs. 
%Even though the broadening of the lines due to turbulence is included, this is not enough to explain the measured broadening in the observations. We find that, in order to explain the measured width, it is necessary to include both, a blue and a red additional absorption components. In particular, we find that we require rather large components with velocities (with respect to the rest-frame of the planet) of --13.5\,\kms\ and +8\,\kms\ which need to be applied to about 40\% and 20\% fractions of the atmosphere along the terminator. 

%Even though we included also that due to turbulence (${\rm v}_{turb}=\sqrt{5kT/(3m)}$, where $k$ is the Boltzmann constant, $T$ is the temperature, and $m$ is the mass of He), they were not enough to explain the measured broadening in the observations and the broadening of the radial velocities of our model along the observer's line of sight, they are not enough to explain the measured broadening in the observations. 3-D model simulations would be required to fully explain the shape of the absorption profile. They can, however, be fitted quite well with two effective red and blue shifted components (with respect to the rest-frame of the planet) of --13.5\,\kms\ and +8\,\kms\  which need to be applied to about 40\% and 20\% fractions of the atmosphere along the terminator. These two components give us a notion of the winds in the terminator of this planet.

The inclusion of the broadening of the lines due to turbulence ($v_{\rm turb} = \sqrt{5kT/3m}$,
where $m$ is the mass of a helium atom), in addition to the standard Doppler broadening, was not enough to explain the measured broadening in the observations (cyan line in Fig.~\ref{absorption}). However, when we included the broadening due to the component of the radial velocities of the gas calculated in our model along the observer's line of sight, (see Eq.\,15 in \citealt{Lampon2020}), then we were able to explain the absorption line width (red curve in Fig.~\ref{absorption}). Because of the weak surface gravity of this planet, the obtained radial velocities were rather large, even at relatively short radii. In particular, we obtained radial velocities in the range of 5 to 20\,km\,s$^{-1}$ for $r$ = 1--10\,$R_{\rm P}$. These velocities, particularly at low radii, induce a rather significant broadening as shown in Fig.~\ref{absorption}.
Nevertheless, we observed that the peak of the absorption was slightly shifted to blue wavelengths, indicating that there may be a net blue wind flowing from the day to the night side, for which we estimated a net velocity shift of --3.2$\pm$1.3\,km\,s$^{-1}$. This result is similar to that  of --1.8\,km\,s$^{-1}$ found by \citet{Alonso2019} for HD\,209458b, which was also interpreted as a net day-to-night thermospheric wind. Our model, being 1D and spherically homogeneous, was not able to predict any net blue or red component. Hence, the calculation shown in  Fig.~\ref{absorption} (red curve) was obtained by imposing a net shift of --3.2\,km\,s$^{-1}$ on the radial velocities computed by our model.

Our 1D hydrodynamic and spherically symmetric model was based on the assumption of a constant sound speed and, hence, it was unable to discriminate among the temperature and the mass-loss rate. That is, both quantities are degenerate. However, it had the advantage of being computationally very efficient, which allowed us to explore a wide range of atmospheric temperatures and mass-loss rates that were compatible with the \het\ absorption. Hence, this measurement significantly constrained the parameter space of those quantities.
We performed calculations by covering a range of maximum temperatures from 6000\,K to 9000\,K and found that the mass-loss rate, \mlr, is confined to a range of 3\,$\times\,10^{10}$\,\gs\ for $T$ = 6000\,K  to about 10\,$\times\,10^{10}$\,\gs\ for $T$ = 9000\,K . 
%These evaporation rates are significantly larger (by one to two orders of magnitude) than those found for \hd20 at lower temperatures  \citep[see, e.g.,][]{Alonso2019} but are comparable or slightly smaller for temperatures close to 11000\,K.

For HD~209458\,b, \citet{Lampon2020} derived mass-loss rates of 1.3\,$\times\,10^{10}$\,\gs\ and 1.3\,$\times\,10^{11}$\,\gs\ for those temperatures (derived for a H/He ratio of 98/2), which are slightly smaller at about 6000\,K but slightly larger at a temperature of 9000\,K than those derived here for GJ 3470\,b for the canonical H/He ratio of 90/10. However, if considering the same H/He ratio, the mass-loss rates are about a factor of 10 larger in GJ~3470\,b than in HD 209458\,b.

The mass-loss rate of GJ 3470\,b derived by \citet{Bou18} was in the range of (1.5--8.5)\,$\times\,10^{10}$\,\gs. The lower limit was derived assuming the mass-loss rate of only neutral hydrogen atoms (that is, neither H$^+$ nor helium were included), while the upper limit was obtained by using the energy-limited approach. Our value at 6000\,K is about twice their lower limit, but both are consistent since they only include neutral hydrogen. Our rate at a temperature of 9000\,K is slightly larger than their upper limit.

%{\color{red} \mlp{Reason for his??? The density of \gj34 is larger (0.76$\rho_J$ versus 0.28$\rho_J$ for \hd20). This should lead to smaller \mlr\ (for a given temperature). The behaviour is probably caused by its smaller size and smaller gravity. Action: Explain the reasons for this!!!}}

%More details on the modelling and on the temperature and mass-loss rates results will be given in a future paper.

%Until now, H$\alpha$ has been only detected in the atmosphere of highly-irradiated exoplanets orbiting early-type stars such as KELT-9b (Yan et al. 2018) and MASCARA-2b (Casasayas-Barris et al. 2018 and 2019), while the He I has been detected in several planets, most of them around active F, G, K-type stars, observing that most of the ones presenting H$\alpha$ that have been studied in the near-infrared region, do not present helium.

\section{Discussion and conclusions}
\label{sec:discussion}

Here we report the detection of He~{\sc i} absorption in the upper atmosphere of GJ~3470\,b. To understand this observation in a broader context, it is important to compare the properties of GJ~3470\,b with two other well-studied Neptune planets: GJ~436\,b \citep{Butler2004,Gillon2007} and \hat11 \citep{Bakos2010}. All three planets have very close radius values (see Table~\ref{Tab:comp}, where the physical properties of all three planets are summarized). GJ~436\,b and \hat11 have also nearly the same mass, density, and age \citep{Demory2013, Fraine2014}, while GJ~3470\,b is less massive and only about half the average bulk density. 

For GJ~436\,b, very significant extra absorption during transit has been observed in \lya. Both \citet{Kulow2014} and \citet{Ehrenreich2015} detected an extended transit with a comet-like tail structure, reaching a depth of almost 50\,\% of the stellar flux. Despite this, absorption in \halpha\ during transit has not been detected, and \citet{Cauley2017} suggested that the large cloud of neutral hydrogen surrounding GJ~436\,b is almost entirely in the ground state. While a strong absorption in \het\ was theoretically predicted by \citet{Oklopcic2018}, \citet{Nortmann2018Science} found no detectable evidence for it.  
%\mlp{Enric: Another point for possible inclusion: 
However, \citet{Salz2016} showed that the concentration of ionized hydrogen in GJ~436\,b is significantly lower than in \gj34 at high altitudes (at radii larger than $\sim$3\rp, the region where according to our model the \het\ is mainly formed). We recall that the major formation process of \het\ is recombination from \hep\ + e$^-$. Thus, a lower density of ionized hydrogen leads to a lower electron concentration and, consequently, to a less efficient \het\ formation, which is in line with the observations of \citet{Nortmann2018Science}.
In the case of HAT-P-11\,b, there are no published detections of extra absorption either in \halpha or \lya, but \citet{Allart2018} detected a strong signature of \het\  absorption during transit. 

For GJ~3470~b, based on ultraviolet observations of the \lya\ absorption, \citet{Bou18} estimated a mass-loss rate of (1.5--8.5)\,$\times\,10^{10}$\,\gs, comparable to that of hot Jupiters, and concluded that the planet could already have lost up to 40\,\% of its mass over its 2 Gyr lifetime. This observation is roughly in line (depending on the actual thermospheric temperature) with the \mlr\ derived from our analysis of the observed \het\ absorption described above. We obtained a value of  (3--10)\,$\times\,10^{10}$\,\gs\ for a temperature range of 6000\,K to 9000\,K. Those values are also comparable to the ones obtained by \citet{Lampon2020} for \het\ \mlr\ of the hot Jupiter \hd20.
We caution, however, that there is a strong dependency of these values on the assumed H/He ratio values, which are currently unknown. \citet{Lampon2020} derived similar mass-loss rate by using H/He = 98/2, imposed by the Ly$\alpha$ measurements, but they had large errors as only the wings of the line were detected. If we used only the \het\ measurements and assumed the same H/He = 90/10 for both \hd20 and GJ~3470\,b, then the mass-loss rate of GJ~3470\,b would be about a factor 10 larger than caculated in Sect.~\ref{sec.modeling}.

%\mlp{MLP: Point for discussion: \citet{Lampon2020} derived similar MLRs because we used H/He=98/2, imposed by the Lyman-alpha measurements. If we were to use ONLY the \het\ measurements and assume the same H/He=90/10 in both HD209 and GJ34, then, clearly, the MLR of GJ is about a factor of 10 larger. We do not know what is (would be) the H/He ratio in GJ. Also, the H derived from Lyman-a for HD209 has a big error (it is detected only in the wings and with large error bars). So, regarding next par., I do not really know what to conclude. To me it seems that evaporation is larger in GJ than in HD209 (i.e. if using the same H/He).}

It is of particular interest to consider why planets with such similar physical properties display very different upper atmospheric escape properties. 
As discussed in \citet{Nortmann2018Science}, the formation of the He~{\sc i} $\lambda$10830\,\AA\ triplet in exoplanet atmospheres is directly linked to the stellar irradiation with $\lambda<$ 504\,\AA, which ionizes the neutral helium atoms, with a subsequent recombination with electrons. Therefore, it is essential to know the X-ray and extreme ultraviolet (XUV) irradiation in this spectral range. The X-ray observations of GJ~3470 reveal a moderately active star ($\log L_{\rm X}/L_{\rm bol}=-4.8$) with some flaring variability (J.~Sanz-Forcada et al. in prep).
The analysis of the X-ray spectrum and ultraviolet lines was used to construct a coronal model and calculate a spectral energy distribution in the full range 1--1200\,\AA\  \citep[][and Sanz-Forcada et al. in prep.]{Bou18}.
The XUV luminosity in the 5--504\,\AA\ range is $L_{\rm XUV~He}=5\times 10^{27}$\,erg\,s$^{-1}$, yielding an irradiation in this band at the distance of GJ~3470\,b of $f_{\rm XUV~He}=1435$\,erg\,s$^{-1}$\,cm$^{-2}$. 
Thus, the $f_{\rm XUV, He}$ of GJ~3470\,b is similar to that of HAT-P-11\,b, but it is almost one order of magnitude larger than that of GJ~436\,b (Table~\ref{Tab:comp}). While the youth and lower density of GJ~3470~b compared to the other two Neptunes surely plays a role, our results suggest that \het\ ionization is mainly driven by XUV stellar irradiation. 

%\mlp{MLP: I agree, although there might be another reason: the XUV stellar ionizing the H (500-1000\,\AA), which will provide the necessary electrons. Maybe we do not want to make this comment?}

As mentioned above, we previously analyzed the \het\ absorption detection in the hot Jupiter \hd20 \citep{Alonso2019,Lampon2020}, another CARMENES target of our He~{\sc i} survey.  We noticed two significant differences. First, the \het\ absorption profile of \gj34 is significantly wider than in \hd20. Secondly, our model showed that the absorption in \gj34 takes place mainly in the outer regions. Both facts suggest that \gj34 has a rather expanded atmosphere with strong winds prevailing in its upper thermosphere. These results are in line with its lower gravity with respect to \hd20. Moreover, we found that the \mlr/T relationship derived from the measured \het\ absorptions are rather different: \gj34 exhibits comparable or even larger \mlr ~(for the same temperature) than the hot Jupiter \hd20. These results suggest that escape of \gj34 is possibly driven by a different process than in \hd20.

\begin{table*}
\centering
\caption{Physical planet parameters for GJ~3470b, HAT-P-11~b, and GJ~436~b.}
\begin{tabular}{lccc}
    \hline
    \hline
    \smallskip
Parameter & GJ~3470~b $^{1-5}$  & HAT-P-11~b $^{5,6}$ & GJ~436~b $^{7-10}$ \\
    \smallskip
Host sp. type                    & M2.0\,V              & K4\,V                & M2.5\,V \\
Radius [\Rjup]              & 0.36$\pm$0.01     & 0.389$\pm$0.005   & 0.374$\pm$0.009 \\
Mass [\Mjup]                & 0.036$\pm$0.002   & 0.0736$\pm$0.0047 & 0.0728$\pm$0.0024\\ 
Density [$\mathrm{g\,cm^{-3}}$]      & 1.036$\pm$0.119 & 1.658$\pm$0.127 & 1.848$\pm$0.163 \\
$T_{\rm eq}$ [K]            & 733$\pm$23        & 832$\pm$10        & 686$\pm$10 \\ 
$F_{\rm EUV}$ [$\mathrm{erg\,s^{-1}\,cm^{-2}}$]     & 1435$\pm$80       & 2109$\pm$153      & 197$\pm$9 \\
Age [Gry]                  & <3                & 6.5$\pm$5.0       & 6.0$\pm$5.0 \\
\het\ Absorption [\%]  &  1.5$\pm$0.3  & 1.08$\pm$0.05 &  <0.41 \\
\\[-1em]
\hline
%\hline
\end{tabular}\\
\begin{list}{}{}
\item {\bf References.} 
$^1$\citet{Biddle2014};
$^2$\citet{Kosiarek2019};
%RL: REFERENCES FOR PLANET PARAMETERS OF GJ3470 USING TABLE 2
%    Radius -> Rp/Rs -> Biddle et al. 2014 (https://ui.adsabs.harvard.edu/abs/2014MNRAS.443.1810B/abstract)
%    Mass -> K, e -> Kosiarek et al. 2019 (https://iopscience.iop.org/article/10.3847/1538-3881/aaf79c/pdf)
%    Mass -> P, i -> Biddle et al. 2014
%    Density -> nothing
%    Teq -> a/Rs -> Dragomir et al. 2015 (https://iopscience.iop.org/article/10.1088/0004-637X/814/2/102/pdf)
$^3$\citet{Dragomir2015};
$^4$This work;
%RL: REFERENCES FOR PLANET PARAMETERS OF HAT-P-11 -> Yee et al. 2018 (https://ui.adsabs.harvard.edu/abs/2018AJ....155..255Y/abstract)
$^5$\citet{Yee2018};
$^6$\citet{Allart2018};
$^7$\citet{2015A&A...577A.128A}
$^8$\citet{Turner2016};
$^9$\citet{Nortmann2018Science};
$^{10}$\citet{Torres2008}.
%RL: REFERENCES FOR PLANET PARAMETERS OF GJ 436 -> Turner et al. 2016 (https://ui.adsabs.harvard.edu/abs/2016MNRAS.459..789T/abstract)
%    Age -> Torres et al. 2008
\end{list}
\label{Tab:comp}
\end{table*}

\begin{acknowledgements}

CARMENES is an instrument for the Centro Astron\'omico Hispano-Alem\'an (CAHA) at Calar Alto (Almer\'{\i}a, Spain), operated jointly by the Junta de Andaluc\'ia and the Instituto de Astrof\'isica de Andaluc\'ia (CSIC).
CARMENES was funded by the German Max-Planck-Gesellschaft (MPG), the Spanish Consejo Superior de Investigaciones Cient\'{\i}ficas (CSIC), the European Union through FEDER/ERF FICTS-2011-02 funds, and the members of the CARMENES Consortium.
We acknowledge financial support from the Agencia Estatal de Investigaci\'on of the Ministerio de Ciencia, Innovaci\'on y Universidades and the European FEDER/ERF funds through projects ESP2016-80435-C2-2-R, ESP2016-76076-R, and BES-2015-074542, and AYA2016-79425-C3-1/2/3-P, % UCM+CAB+IAA 
the Deutsche Forschungsgemeinschaft through the Research Unit FOR2544 ``Blue Planets around Red Stars'' and the Priority Program SPP 1992 ``Exploring the Diversity of Extrasolar Planets'' RE 1664/16-1,
the National Natural Science Foundation of China through grants 11503088, 11573073, and 11573075,
and the Natural Science Foundation of Jiangsu Province through grant BK20190110.
Finally, we wish to thank Dr. Vincent Bourrier and an anonymous referee for discussion and comments that helped to improve the contents of this manuscript.

\end{acknowledgements}

\bibliographystyle{aa.bst} 
\bibliography{biblio.bib}

%\appendix

%\section{Contaminated transit model and candidate validation}
%\label{sec:transit_model}

%The candidate validation is based on the \emph{true radius} estimate obtained modelling the multicolour transit photometry with a transit model that includes a physically-based light contamination component parametrised by the effective temperatures of the host and the contaminant (\teffh and \teffc, respectively), and the amount of contamination in some reference passband, \pcref. 

\end{document}